\documentclass[preprint,amsmath,amssymb,floats]{revtex4}
\usepackage{graphicx}
\usepackage{dcolumn}
\usepackage{bm}
\def\Re{\textrm{Re}}
\def\Im{\textrm{Im}}

\def\vk{{\bf k}}
\def\be{\begin{equation}}
\def\ee{\end{equation}}
\def\bea{\begin{eqnarray}}
\def\eea{\end{eqnarray}}

\begin{document}
\title{Superconductivity in the Cuprates: Deduction of Mechanism for D-Wave Pairing Through Analysis of ARPES}
\author{Han-Yong Choi$^1$, C.M. Varma$^2$, Xingjiang Zhou$^3$}
\affiliation{$^1$Department of Physics and Institute for Basic Science Research, 
Sung Kyun Kwan University, Suwon 440-746, Korea. \\
$^2$Department of Physics and Astronomy, University of
California, Riverside, California 92521.\\
$^3$National Lab for Superconductivity, Institute of Physics, Chinese Academy of Sciences, Beijing 100190, China.}

\begin{abstract}
In the 
Eliashberg integral equations for d-wave superconductivity, two different functions $(\alpha^2 F)_n(\omega, \theta)$ and $(\alpha^2 F)_{p,d}(\omega)$  determine, respectively, the ``normal" self-energy and the ``pairing" self-energy. $\omega$ is the frequency of fluctuations scattering the fermions whose momentum is near the Fermi-surface and makes an angle $\theta$ to a chosen axis. 
We present a quantitative analysis of the high-resolution laser based ARPES data on a slightly under doped cuprate compound Bi-2212 and use the Eliashberg equations to deduce the $\omega$ and $\theta$ dependence of $(\alpha^2 F)_n(\omega, \theta)$ for $T$ just above $T_c$ and below $T_c$. Besides its detailed $\omega$ dependence, we find the remarkable result that this function is nearly independent of $\theta$ between the ($\pi,\pi$)-direction and 25 degrees from it, except for the dependence of the cut-off energy on $\theta$.  Assuming that the same fluctuations determine both the normal and the pairing self-energy, we ask what theories give the function $(\alpha^2 F)_{p,d}(\omega)$ required for the d-wave pairing instability at high temperatures as well as the deduced $(\alpha^2 F)_n(\theta, \omega)$. We show that the deduced $(\alpha^2 F)_n(\theta, \omega)$ can only be obtained from Antiferromagnetic (AFM) fluctuations if their correlation length is smaller than a lattice constant. Using $(\alpha^2 F)_{p,d}(\omega)$ consistent with such a correlation length and the symmetry of matrix-elements scattering fermions off AFM fluctuations, we calculate $T_c$ an show that AFM fluctuations are excluded as the pairing mechanism for d-wave superconductivity in cuprates. We also consider the quantum-critical fluctuations derived microscopically as the fluctuations of the observed loop-current order discovered in the under-doped cuprates, and which lead to the marginal fermi-liquid properties in the normal state. We show that their frequency dependence and the momentum dependence of their matrix-elements to scatter fermions are consistent with the $\theta$ and $\omega$ dependence of the deduced $(\alpha^2 F)_n(\omega, \theta)$. The pairing kernel $(\alpha^2 F)_{p,d}(\omega)$ calculated using the experimental values in the Eliashberg equation gives $d-wave$ instability at $T_c$ comparable to the experiments.
 \end{abstract}
\maketitle

\section{Introduction}

The family of Cuprates have the highest superconducting transition temperatures $T_c$ discovered so far. Superconductive pairing is in the "d-wave" symmetry.  Superconductivity with d-wave symmetry and at such high temperatures requires a mechanism different from effective  electron-electron attraction through virtual exchange of phonons.
 
We begin by reviewing the mechanisms for d-wave pairing by a general symmetry analysis of the momentum dependence of the  pairing vertex. We also define two quantities which appear in the analysis of experimental data in the second part and which determine the "normal" and the "pairing" self-energies. We then use experimental data from high resolution laser ARPES at different angles across the fermi-surface to deduce aspects of the momentum and frequency dependence of the ``normal" self-energy both in the normal and the superconducting  states. We summarize the results of the inversion of the Eliashberg Equations to deduce the momentum and frequency dependence of the particle-hole fluctuations which lead to the observed self-energy. There have been innumerable ideas and calculations proposed to understand the properties of the Cuprates. We consider two specific mechanisms which have been proposed for d-wave pairing due to a purely electronic mechanism: Exchange of Antiferromagnetic fluctuations which are prominent in very underdoped cuprates and of quantum-critical fluctuations from loop current order observed in under doped cuprates in the pseudo-gap region of the phase diagram. We ask, given the experimentally deduced fluctuations and their coupling to fermions, whether one or both is consistent with the measured ARPES spectrum and the measured $T_c$.

Soon after BCS theory, Eliashberg \cite{eliashberg} used the field theory methodology developed for superconductivity by Gorkov \cite{gorkov-sc} to formulate the theory of superconductivity to include the frequency dependence of the effective interactions through exchange of phonons.  
The unambiguous experimental proof that the superconductivity in metals such as $Pb, Sn$, etc. is induced by electron-phonon interaction is given by the analysis of tunneling spectrum by Rowell and McMillan \cite{mcmillan-rowell, 
schrieffer-scalapino} in these metals using the Eliashberg theory and the measurement of the spectrum of phonons by neutron scattering. The theory also provides experimental proofs of its limit of validity. With one important modification, which does not affect the linearized Eliashberg theory which is enough to determine $T_c$,  the theory can be used for pairing in any symmetry of degenerate fermions due to exchange of any kind of fluctuations, provided the conditions for limit of its validity are satisfied. 

In practice, the procedure for extracting and using information for d-wave superconductors and metals, such as the cuprates, in which the high frequency cut-off of the energy of relevant fluctuations is an order of magnitude larger than that of lattice vibrations is much more demanding of both data and analysis.
Even for phonons, Rowell and McMillan used data which had a relative accuracy in the measured conductance of $0.2\%$ over a few times the highest phonon frequency and from above $T_c$ to well below $T_c$. This led to completely reliable conclusions. For reasons that have  been discussed \cite{vekhter-cmv}, ordinary tunneling or STM is not suitable for deducing spectrum of fluctuations promoting anisotropic pairing as in the cuprates. For that we must turn to ARPES and measurements and analysis at various angles across the Fermi-surface. The data needs to be consistent and reliable from above $T_c$, to at least half of  $T_c$, at various angles and over an energy range of about $0.5$ eV, which we will show is the upper cut-off of the fluctuations. We have relied on the best available laser -ARPES data, from the group of one of us (Xingjiang Zhou at Beijing). Even this data at present is only reliable in the superconducting state to about $1\%$ up to only an energy of about 0.2 eV. The data above $T_c$ at angles from the diagonal to the BZ to about 25 degrees to it is reliable to this accuracy to about 0.5 eV. We expect future data to completely solve this problem but on the basis of existing data some fairly reliable conclusions can be drawn.

\section{Pairing Symmetry}

We will show in this section that  in the spin-singlet channel, s-wave pairing is favored when the scattering of fermions from ${\bf k}$ to ${\bf k}'$ with both near the fermi-surface is independent of the angle between ${\bf k}$ and ${\bf k}'$ and d-wave pairing is favored when the strongest scattering is between
${\bf k}$ and ${\bf k}'$  oriented $\pi/2$ with respect to each other. $\pi/2$ scattering can occur both through the well known case of the exchange of Antiferromagnetic fluctuations, provided they are sharply enough peaked near a commensurate wave-vector which spans the fermi-surface, or through exchange of current fluctuations which we will specify below.

These facts about the favored pairing symmetry are implicit in earlier work \cite{miyake} and can (in most cases) be deduced from the momentum and spin-dependence of the effective interaction Hamiltonian, written in terms of the irreducible interaction function $I_{S}({\bf k},{\bf k+q}, \omega)$ with momenta ${\bf k}, {\bf k+q}$ at the fermi-surface, total spin $S$ and energy transfer $\omega \to 0$. 

The pairing vertex $ I_{S}({k, k+q},\omega)$ scattering fermions at $({\bf k}, \alpha), (-{\bf k}, \beta)$ to $({\bf k+q}, \gamma), ({-\bf k +q}, \delta)$ with an energy transfer $\omega$ may be written as
\be
\label{irr-int}
I_{S}({k, k+q}, \omega) \equiv g_{\alpha,\beta}({\bf k, k+q})g_{\gamma,\delta}({\bf -k, -k-q}) \mathcal{F}_{\alpha,\beta,\gamma,\delta}({\bf q, k},\omega).
\ee
$\mathcal{F}_{\alpha,\beta,\gamma,\delta}({\bf q}, {\bf k}, \omega)$ is the propagator of the fluctuations which are exchanged by the fermions,  and $g_{\alpha,\beta}({\bf k, k+q})$ is the scattering matrix.

An interaction of the spin-rotational invariant form may be separated into spin-independent and spin-dependent parts:
\be
\label{modelSF}
I_{\alpha,\beta,\gamma,\delta}({\bf k},{\bf k'}) = 1/2 \sum_{{\bf kk'}} \Big(I_1({\bf k,k'}) \delta_{\alpha \beta} \delta_{\gamma \delta} + I_2({\bf k,k'}) \sigma_{\alpha \beta}\cdot \sigma_{\gamma \delta}\Big) c^+_{{\bf k}, \alpha} c^+_{{\bf -k}, \gamma} c_{{\bf -k'}, \delta} c_{{\bf k'}, \beta}.
\ee

The interaction in the spin singlet $(S=0)$ channel must be even under the interchange ${\bf k} \to {\bf k'}$ and must be odd under this interchange for the triplet $(S=1)$ channel. Therefore both $I_1$ and $I_2$ can contribute to the $S=0$ channel, but only $I_2$ can contribute to the $(S=1)$ channel. 
The matrix element for pairing in the $S=0$ and $S=1$ channels after the appropriate spin-traces are
\bea
\label{sing.trip}
I(S=0, {\bf k},{\bf k}') &=&\frac{1}{2} [I_1({\bf k},{\bf k}') - 3(I_2({\bf k},{\bf k}') + I_2({\bf -k},{\bf -k}'))], \\
I(S=1, {\bf k},{\bf k}') &=& \frac{1}{2} [I_2({\bf k},{\bf k}') - I_2({\bf -k},{\bf -k}')].
\eea
Consider short-range interactions in real space so that they can be written as separable functions of ${\bf k}$ and ${\bf k'}$. The separable functions  decompose into sums over different angular momentums $\ell$ if the Fermi-surface is isotropic or more generally into irreducible representations of the point-group of the lattice. For pairing on a single Fermi-surface, the physics to discern the symmetry of superconductivity can be learnt from considering an isotropic fermi-surface with both ${\bf k}$ and ${\bf k'}$ on the fermi-surface. In this case case both $I_1$ and $I_2$ have the form 
$\propto \sum_{\ell} a_{\ell} (k_F) P_{\ell}(\hat{k})P_{\ell}(\hat{k'})$.

The projected pairing interaction in the $\ell$-th angular momentum channel for $S=0$ from Eq. (\ref{modelSF}, \ref{sing.trip}) is usefully written
as an integral over the momentum transfer $q = |{\bf k}-{\bf k'}| \approx 2 k_F \sin (\theta/2)$, where $\cos \theta = \hat{\bf k_F}\cdot \hat{\bf k'_F}$. Then with $\sin \theta/2 = x$ and for $\omega \to 0$,
\bea
\label{singlet}
 I(S=0, \ell) = 2 \int_0^1 dx x P_{\ell}(1-2x^2)[
I_1 (2k_Fx) - 3 I_2(2k_F x)], ~~ {\ell}~ even.
\eea
Several simple and important points may also be noted from Eq. (\ref{singlet}):\\
 (i) An interaction independent of x, i.e a  momentum independent interaction  gives 0 for all  $\ell$, except $\ell =0$. This represents the fact that such an interaction is a $\delta$- function  in real space and any finite $\ell$ pair wave-function has zero amplitude at the origin.\\
 (ii) An attractive interaction $I_1 <0$ is required for the $\ell =0$ case but $I_1 <0$ provides a repulsive interaction for $\ell \ne 0$. \\
 (iii) An antiferromagnetic $I_2$ interaction is repulsive for $\ell =0$. It is attractive for $\ell = 2$ only if it peaks sufficiently near the zone boundary. \\
 (iv) For the contribution both from $I_1$ and from $I_2$,  $\ell =2$ is favored when the strongest interaction is at $x^2 = \pm 1/2$, i.e. that the initial and final states are at $\pi/2$ with respect to each other. 

It is straightforward to extend this analysis \cite{miyake} to the more realistic case for the momentum dependence of the interaction taking the crystal symmetry into account.  We refer to the original paper for  several different crystal symmetries. For a square lattice, of relevance to the cuprates, the peaking of AFM fluctuations near the $(\pi,\pi)$ point and doping not too far from half-filling, strongest scattering of fermions near the Fermi-surface occurs through an angle $\pi/2$ leading, as above, to d-wave pairing. At the same time, the ``normal" self-energy of the fermions depends on their direction of the antiferromagnetic vector ${\bf Q}$ with respect to the crystalline axes. The strength of both effects depends on how steep is the increase in $I({\bf q})$ near the zone-boundary and on the details of the electronic dispersion $\epsilon({\bf k})$ near the chemical potential. 

While the arguments above  provide the basic physical principles for discussing pairing symmetry, it is necessary also to consider the frequency dependence of the pairing vertex for a theory of $T_c$.  The considerations determining $T_c$ have been discussed elsewhere \cite{cmv-Tcreview}; here we will decipher the spectrum directly from the experiments. 

In order to obtain quantitative conclusions from analysis of ARPES data, it is useful to define two functions. From Eq. (\ref{irr-int}) and the ensuing discussion, we define a function which enters as the Kernel for Eliashberg Equation for the ``pairing" or ``anomalous" self-energy in the d-wave channel:
\bea
\label{a2Fa}
(\alpha^2F)_{p,d}(\omega) &\equiv& \int \frac{dS({\bf \hat{k}}_F)}{|v({\bf \hat{k}}_F)|} \int \frac{dS({\bf \hat{k}'}_F)}{|v({\bf \hat{k}'}_F)|}  
 \Big( g_1({\bf k}, {\bf k}') g_1(-{\bf k}, -{\bf k}') Im \mathcal{F}_1({\bf k},{\bf k}', \omega) \\ \nonumber  & - &3 g_2({\bf k}, {\bf k}') g_2(-{\bf k}, -{\bf k}') Im \mathcal{F}_2({\bf k},{\bf k}', \omega) \Big) P_{2}({\bf \hat{k}}) P_{2}({\bf \hat{k}'})/ \int \frac{dS({\bf \hat{k}'}_F)}{|v({\bf \hat{k}'}_F)|}.
\eea
Here $g_1,  \mathcal{F}_1$ and $g_2,  \mathcal{F}_2$ are the spin-independent and spin-dependent vertices and Fluctuation propagators, in complete analogy with Eq. (\ref{modelSF}). 
 $P_{2}({\bf \hat{k}})$ is the appropriate projection operator for the given crystal symmetry, ($(1/\sqrt{2})\cos (2 \theta)$ for isotropic case), and the integrations are over the Fermi-surface. 
 
Equally necessary to define is the Kernel for the equation for the ``normal" self-energy in which the vertices and the spin-sums enter differently and which explicitly depends on the direction on the Fermi-surface:
\bea
\label{a2Fn}
(\alpha^2F)_{n}({\bf k}, \omega) &\equiv &  \int \frac{dS({\bf \hat{k}}'_F)}{|v({\bf \hat{k}}'_F)|}  \Big(|g_1({\bf k}, {\bf k}')|^2 Im \mathcal{F}_1({\bf k},{\bf k}', \omega) + 3 |g_2({\bf k}, {\bf k}')|^2 Im \mathcal{F}_2({\bf k},{\bf k}', \omega)\Big).
\eea

Note the absolute magnitude signs in (\ref{a2Fn}) compared to (\ref{a2Fa}), because the ``normal" self-energy must have a pair of conjugate vertices, and the $+3$, rather than $-3$ due to spin-trace in the particle-hole channel. For s-wave superconductivity, the same function is the kernel for both the normal and the pairing self-energies.  

In approximate solutions of the Eliashberg Equations for d-wave pairing, one can define
\bea
\label{lambdas}
\lambda_n &=& \int d\omega \frac{1}{2\pi}\int d\theta \frac{(\alpha^2 F(\omega, \theta))_n}{\omega} \\
\lambda_{a,d} &=& \int d\omega \frac{(\alpha^2 F(\omega))_{p,d}}{\omega},
\eea
Here 
\bea
(\alpha^2 F(\omega, \theta))_n = N(\theta) (\alpha^2F)_{n}({\bf k}_F, \omega),
\eea
where $N(\theta) = (|{\bf v}_F(\theta)|)^{-1} d {\bf k}_t/d\theta$ and ${\bf k}_t$ is the tangent to the fermi-surface.
In terms of $\lambda_n, \lambda_d$, an approximate solution of the Eliashberg equation yields
\bea
T_c \approx \omega_c \exp^{(-\frac{1+\lambda_n}{\lambda_{p,d}})}.
\eea
Better calculations can only be done by numerical solutions but Eq.(\ref{lambdas}) sufficiently explain one of the principle points of this paper.

\section{Deduced Results from  ARPES Experiments}

\begin{figure}
\includegraphics[scale=0.3]{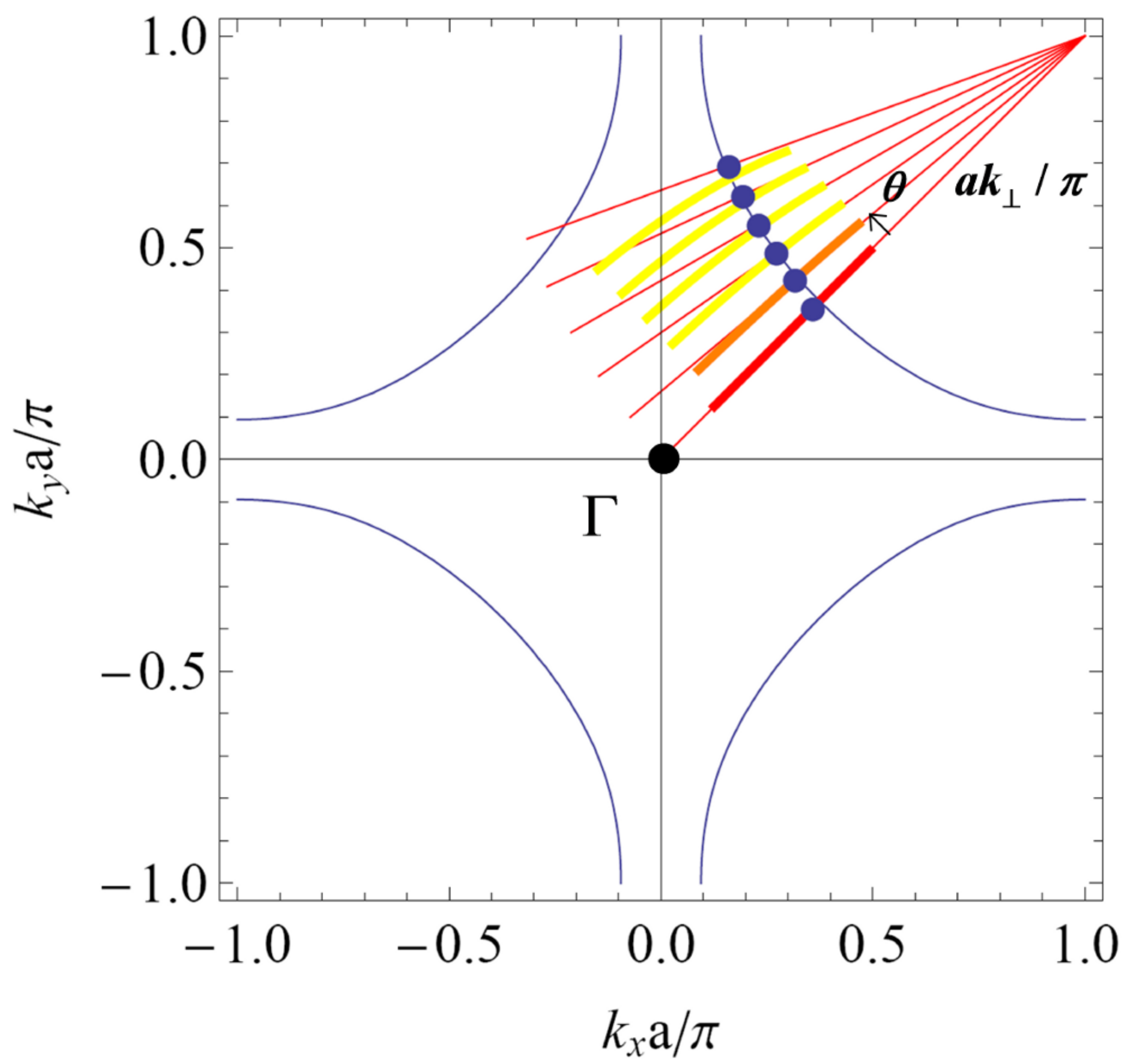}
\caption{The Fermi surface of Bi2212 in the first Brillouin zone.
The blue solid curve is a calculated Fermi-surface and the
solid dots are experimentally determined FS at $\theta=0$, 5, 10,
15, 20, and 25 degrees. $k_\perp$ is the distance from the
$(\pi,\pi)$ point. The thick bars along each cut indicate the
ranges of experimentally measured ARPES MDC data.} \label{fig:FS}
\end{figure}

In three recent papers \cite{choi1, choi2, zhou1}, our collaborators and we have presented a fit to the Laser ARPES data on a slightly underdoped sample of 
BISCCO 2212 with $T_c = 89 K$ and a pseudogap temperature of about $125 K$. We refer to these papers for the detailed procedure for inverting ARPES ( also given in the nodal direction in Ref. (\onlinecite{Carbotte-inversion}) in Cuprates.  Here we only provide a summary of the results.

\begin{figure}
\includegraphics[scale=0.4]{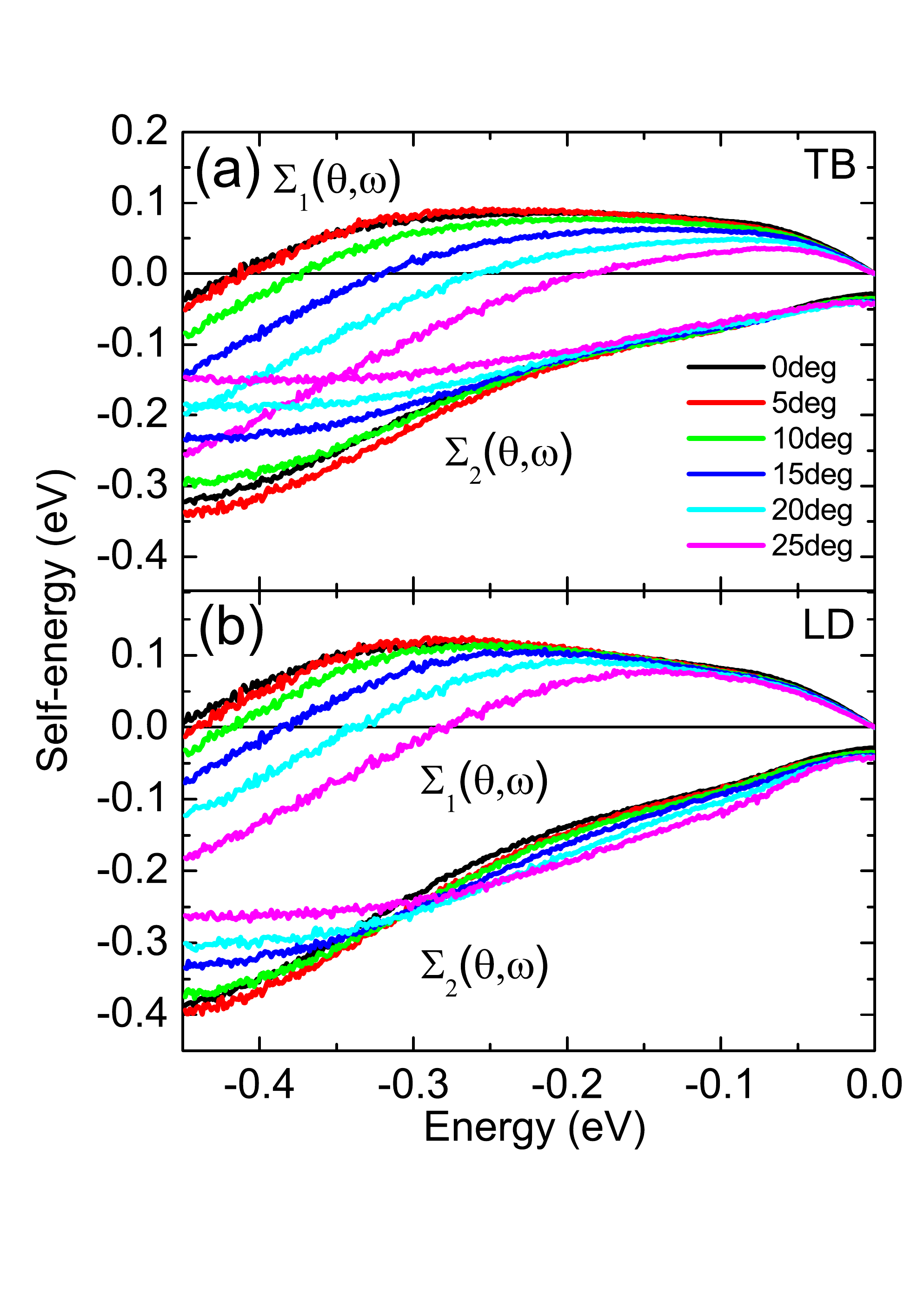}
\caption{The real $Re \Sigma$ and  imaginary $Im \Sigma$ part of self-energy at $T = 107 K$ for the tilt angles $\theta$ = 0, 5, 10, 15, 20, and 25 degrees with respect to the diagonal to the BZ and as a function of positive energy $\omega$. Plot (a) is the result from a detailed fit to band-structure over the whole range of energy while (b) is given to compare the results if a linear 
extrapolation of the band-structure from that near the fermi-energy is adopted.} 
\label{self-energy}
\end{figure}

Data was available along the momentum cuts shown in Fig. (\ref{fig:FS}). We were able to deduce the momentum and frequency dependence of the normal state self-energy at $T= 107 K$ with high accuracy and consistency. The results are shown in Fig. (\ref{self-energy}). The self-energy is labelled by the angle $\theta$ with respect to the diagonal of the BZ and by the energy $\omega$. No perceptible dependence on $|{\vk}-{\vk}_F|$ could be found, except through the upper cut-off in frequency of the spectrum, as described below.

\begin{figure}
\includegraphics[scale=1.5]{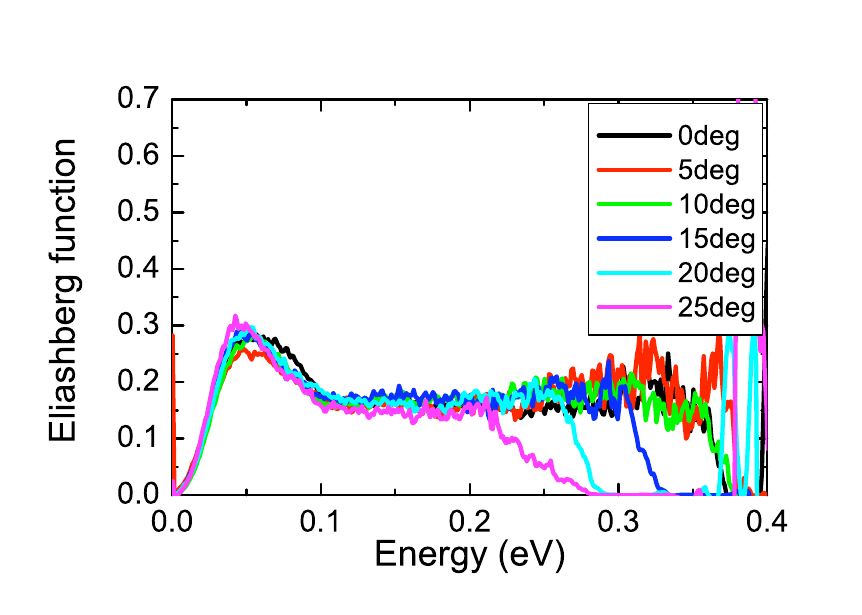}
\caption{The Eliashberg function $(\alpha^2 F)_n(\theta, \omega)$ discussed in the text extracted from the self-energy shown in Fig.(\ref{self-energy}). Notice that despite the considerable dependence of the self-energy on angle $\theta$, the deduced Eliashberg function collapses within the accuracy of its deduction to as single curve below about $0.2 eV$. Above this energy it cuts off, with the cut-off depending on the angle. This cut-off is the same, about 0.4 eV for $\theta = 0, 5, 10$ degrees where the bottom of the band is below the fermi-energy by larger than 0.4 eV. For $\theta = 15, 20, 25$ degrees, the cut-off is simply given by the measured bottom of the band with respect to the fermi-energy at such angles.} 
\label{normal F}
\end{figure}

\subsection{Normal State}
Using the deduced self-energy, we have inverted the Eliashberg equation for the normal state self-energy to deduce the generalized Eliashberg function $(\alpha^2 F (\theta, \epsilon))_n$ defined in Eq. (\ref{a2Fn}). The results are shown in Fig.(\ref{normal F}).

The deduced $(\alpha^2 F (\theta, \epsilon))_n$ is quite striking in two respects:

(1) Even though the self-energy, Fig. (\ref{self-energy}) shows significant angle dependence, the deduced $(\alpha^2 F (\theta, \epsilon))_n$ is  independent of angle to an accuracy of about 10\% below an energy of about 0.2 eV. Above this energy there is an angle dependent cut-off $\omega_c (\theta)$ for $\theta =$ 20 and 30 degrees but the cut-off is independent of angle once it reaches its  maximum value of about 0.4 eV, as for $\theta$ = 15, 10 and 0 degrees. As has been shown (\cite{choi1}), the cut-off is at the bottom of the band with respect to the fermi-energy at any angle except when this value increases beyond about 0.4 eV, where it stays at 0.4 eV.
It follows that the spectrum of the fluctuations has an intrinsic cut-off which is about 0.4 eV.

It also follows from the general relation between the self-energy and this fluctuations spectra and the bare single-particle Green's function that the angle dependence of the self-energy is almost completely due to the dispersion of the bare band.

(ii) As a function of $\epsilon$, $(\alpha^2 F (\theta, \epsilon))_n$ may be considered to be a sum of two features, a nearly $\theta$-independent bump centered at about 0.05 eV and the nearly constant part whose intrinsic cut-off, as discussed, is about 0.4 eV. With that value of the cut-off, the 0.05 feature has about $10\%$ of the total spectral weight, increasing to about $20\%$ at a cut-off of 0.2 eV.

\begin{figure}[tbh]
\centering
\includegraphics[width=0.7\textwidth]{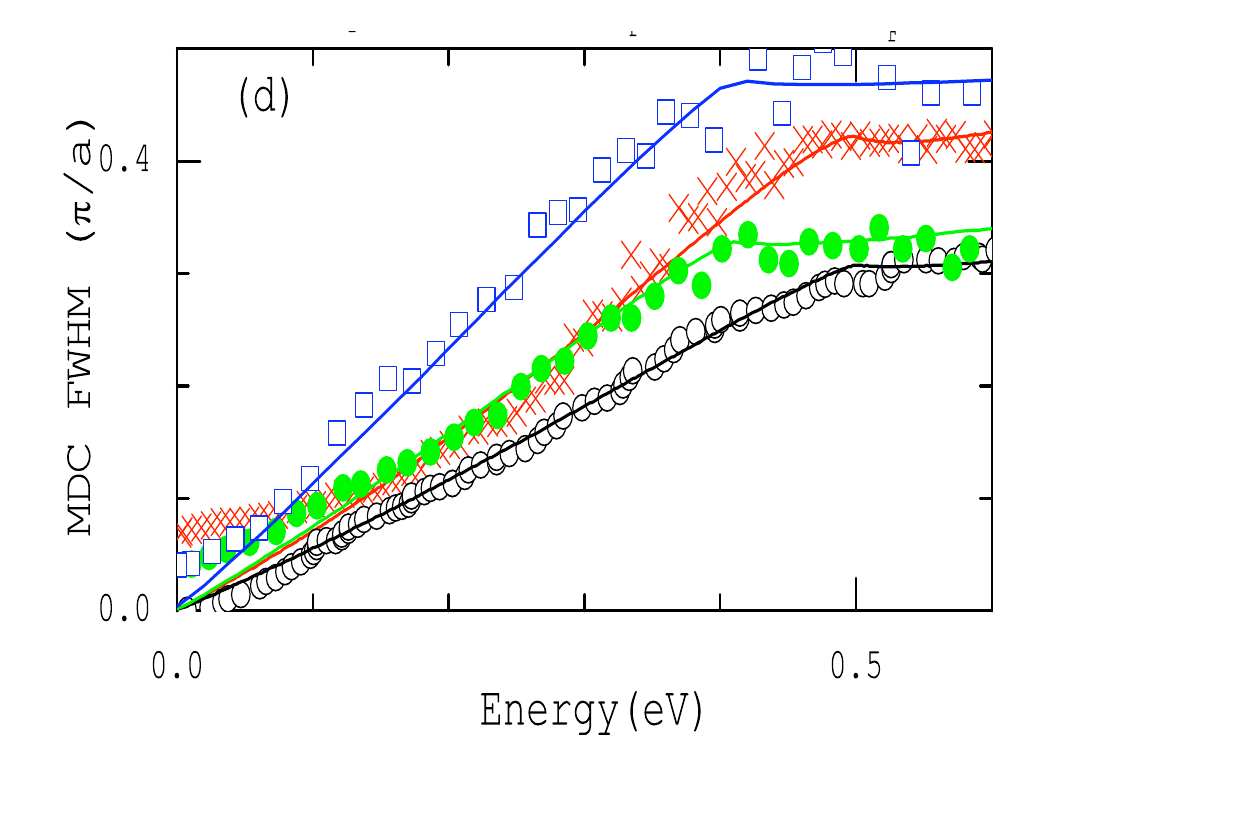}
\caption{The linewidths of the Momentum Distribution curves for different cuprates as a function of the energy near optimal doping and along the diagonal to the BZ. The imaginary part of the self-energy is obtained by multiplying this linewidth with the bare fermi-velocity. The detailed references for each cuprate are given in Ref.(\onlinecite{lijun-prl08})}
 \label{mdcwidth}
\end{figure}

The deductions discussed above are consistent with the normal state self-energy deduced at $\theta =0$ degrees for all the Cuprates by various different groups for which ARPES results have been obtained. See fig. (\ref{mdcwidth}).  The imaginary part of the self-energy for all of them is linear in energy (above an energy which depends on temperature or pseudo gap energy) and has a distinct energy above which it is constant. This is  consistent with scattering from a nearly constant in energy spectrum with a well-defined cut-off. The deduced $(\alpha^2 F (\theta, \epsilon))_n$ at $\theta =0$ is quantitatively consistent with an earlier deduction \cite{Carbotte-inversion} from ARPES spectrum in the same compound, and also qualitatively consistent with the deduction from optical conductivity spectrum \cite{Carbotte-optics},  \cite{vandermal-optics} which preferentially weights the nodal quasi-particles because of their larger fermi-velocity.

\subsection{Superconducting State}


As explained above and elsewhere \cite{vekhter-cmv, choi1,choi2}, unlike the Eliashberg Equations for "s-wave" pairing, it is necessary in the superconducting state for "d-wave" pairing to extract separately the Eliashberg function  $(\alpha^2 F)_n$, which occurs as the kernel in the integral equation for the "normal" self-energy, and the Eliashberg function $(\alpha^2 F)_a$ which occurs as the kernel in the integral equation for the "anomalous" or pairing self-energy. The "normal" self-energy has to maintain the same sign around the Fermi-surface and therefore is proportional in an isotropic system to the form $a_0 + a_4 \cos^2(2\theta)+..$, or equivalent harmonics in a lattice symmetry, while the leading terms for the anomalous self-energy is proportional to $b_0 \cos(2\theta) + b_2 \cos(6\theta) +...$. To our knowledge, this separation can only be accomplished only from analysis of Angle-Resolved Photoemission data. 

The information on the pairing self-energy is contained only in the difference in the ARPES spectra in the superconducting state and the normal state. This difference is expected \cite{choi2} to be less than 1$\%$ above an energy of a few times the superconducting gap. Above such energy, the noise in the data is at present significantly larger than 1$\%$. Therefore, we have not been able to extract the pairing self-energy and to directly deduce $(\alpha^2 F)_{p,d}(\omega)$ from the data over the fully energy range of 0.4 eV. Below about $0.1$ eV, where we can deduce it quite well, it essentially follows the angle-independent deduced $(\alpha^2 F)_n(\theta, \omega)$. 
Indeed, we do not know of any proposed or reasonable model in which the two functions have quite different frequency dependence in the energy range well above the superconducting gap. But it would be ideal to have data which is about 1/2 an order of magnitude better to completely settle the shape of $(\alpha^2 F(\theta, \epsilon))_a$.

The deduced $(\alpha^2 F)_n(\theta, \omega)$ in the superconducting state is noisier than its deduction in the normal state, especially for larger $\theta$ and higher energy. Please see Ref(\onlinecite{choi2}) for the results at various temperatures and angles.

The principal conclusion are that at $\theta = 0$ degrees, the fluctuations below $T_c$, within the uncertainty of determination, are identical to the fluctuations above $T_c$. For larger angles and for energies above about $0.1$ eV, they are similarly identical to the fluctuations above $T_c$. There is a growth of the intensity in the hump around $50$ meV for lower temperatures and larger angles, and a possible growth of a new peak at about $15$ meV. These features are probably related to the loss of dissipation due to the opening of the superconducting gap and would be interesting to study theoretically in greater detail. However, since they cannot affect $T_c$, we will not dwell on them further here. So, in effect we are led only to discuss the bulk of the spectra which does not change from below $T_c$ to above $T_c$.

The limit of validity of the Eliashberg theory for the range of parameters we have uncovered in the Cuprates has been analytically discussed in Ref.(\onlinecite{choi1}). But the conclusive answer can only come from analysis of further experiments as we will discuss near the end of the next section.

\section{Discussion of Results and Conclusions}

In this section, we discuss the implications of the finding that
the Eliashberg function $(\alpha^2F(\theta, \omega))_n$  defined in Eq.
(\ref{a2Fn}) is nearly independent of $\theta$ in the normal state and just below $T_c$.   

It should be recognized {\it a priori} that the superconducting $T_c$ is the property of the normal states. Indeed the normal state self-energy just above $T_c$ includes all forms of scattering, both spin-dependent as well as spin-independent and scattering from all initial momenta ${\bf k}$ to all final momenta ${\bf k'}$. It is unlikely that fluctuations invisible in the normal state determine $T_c$; at least no such idea has been expounded. Given this the crucial issue to address is how fluctuations which lead to a nearly $\theta$-independent $(\alpha^2F(\theta, \omega))_n$ are reconcilable with the same fluctuations promoting $d$-wave pairing, i.e. also giving large values of $(\alpha^2F(\omega))_{a,d}$, defined by Eq. (\ref{a2Fa}).

We discuss two specific ideas for the source of d-wave superconductivity, (i) Coupling of AFM fluctuations to fermions \cite{msv}, \cite{scala-loh}, \cite{Scalapino-physreports} , \cite{Millis90prb} and (ii) coupling of quantum-critical loop current fluctuations to fermions. 

\subsection{AFM fluctuations}

In this section, we show that given the deduced $(\alpha^2F(\theta, \omega))_n$, AFM fluctuations can be excluded as a source of pairing interactions. 
An important point to note for this and the ensuing discussion is that it follows from Eqs. (\ref{a2Fa}, \ref{a2Fn}) that the pairing vertex $I$ depends both on the matrix elements $g_{\alpha,\beta}({\bf k, k+q})g_{\gamma,\delta}({\bf -k, -k-q})$  and the spectra of the fluctuations $\mathcal{F}_{\alpha,\beta,\gamma,\delta; \nu}({\bf q, k},\omega_n)$.
The idea of $\ell = 2$ pairing through spin-interactions $I_2$ \cite{msv}, \cite{scala-loh}, \cite{Scalapino-physreports} is quite well developed. For the Hubbard model, the matrix element for scattering the fluctuations, $g_{\alpha,\beta}({\bf k, k+q})g_{\gamma,\delta}({\bf -k, -k-q})$ is momentum independent, being simply the square of the local interaction energy $U^2$ \cite{Scalapino-physreports, msv} in the weak-coupling limit. At strong coupling the matrix element can acquire in addition the symmetry of the exchange energy. In neither case can it provide the dominant $\pi/2$ scattering discussed above required for d-wave pairing. On the other hand, for a square lattice with AFM fluctuations strongly enough peaked at nearly commensurate ${\bf q}$, such a scattering is provided by $\mathcal{F}$ as discussed in Sec. II and elsewhere \cite{Scalapino-physreports} .  

Therefore if the fluctuations scattering the fermions are dominantly AFM fluctuation, we need to deduce $\mathcal{F}_{\alpha,\beta,\gamma,\delta; \nu}({\bf q, k},\omega_n)$ of an AFM form consistent with the experimentally deduced
 $(\alpha^2F(\theta, \omega))_n$. From this procedure we can deduce the parameters in the commonly assumed form of antiferromagnetic (AF) fluctuations.  A phenomenological form 
for them may be written
as\cite{Millis90prb,Chubukov03}

 \be
\label{eq:chiaf} 
Im {\mathcal F}_{AF}({\vk},\vk',\omega)= \frac{m_0^2\xi^2 ~
\omega/\omega_{AF}}{(({\vk-\vk'}-{\bf Q})^2\xi^2 +1
)^2+(\omega/\omega_{AF})^2}. 
 \ee
 
\begin{figure}
\includegraphics[scale=0.7]{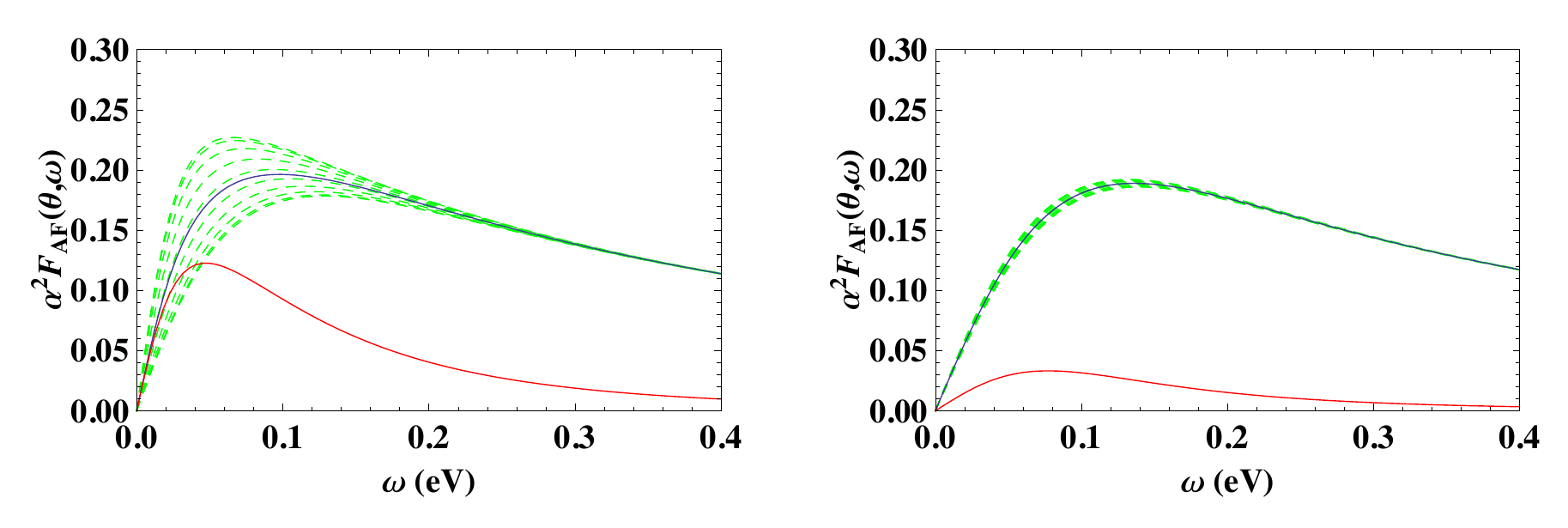}
\caption{The functions discussed in the text for $\omega_{AF} = 0.03 eV$. The left panel is for correlation length
$\xi/a =1$ and the right for $\xi/a = 1/\pi$. The dashed curves are for from top to bottom from the antipodal direction to the nodal directions by  increase of 5 degrees. For $\xi/a = 1/\pi$, the curves are nearly independent of $\theta$. The red curve which determines $\lambda_d$ is $\alpha^2 F_{p,d}(\omega)$  and the blue curve is the s-wave average that determines $\lambda_n$. The $T_c$ calculated for $\xi/a =1/\pi$ and $\xi/a =1$ from the solution of the Eliashberg equations are less than 0.1 K and 12 K respectively.}
\label{AF1}
\end{figure}

\begin{figure}
\includegraphics[scale=0.7]{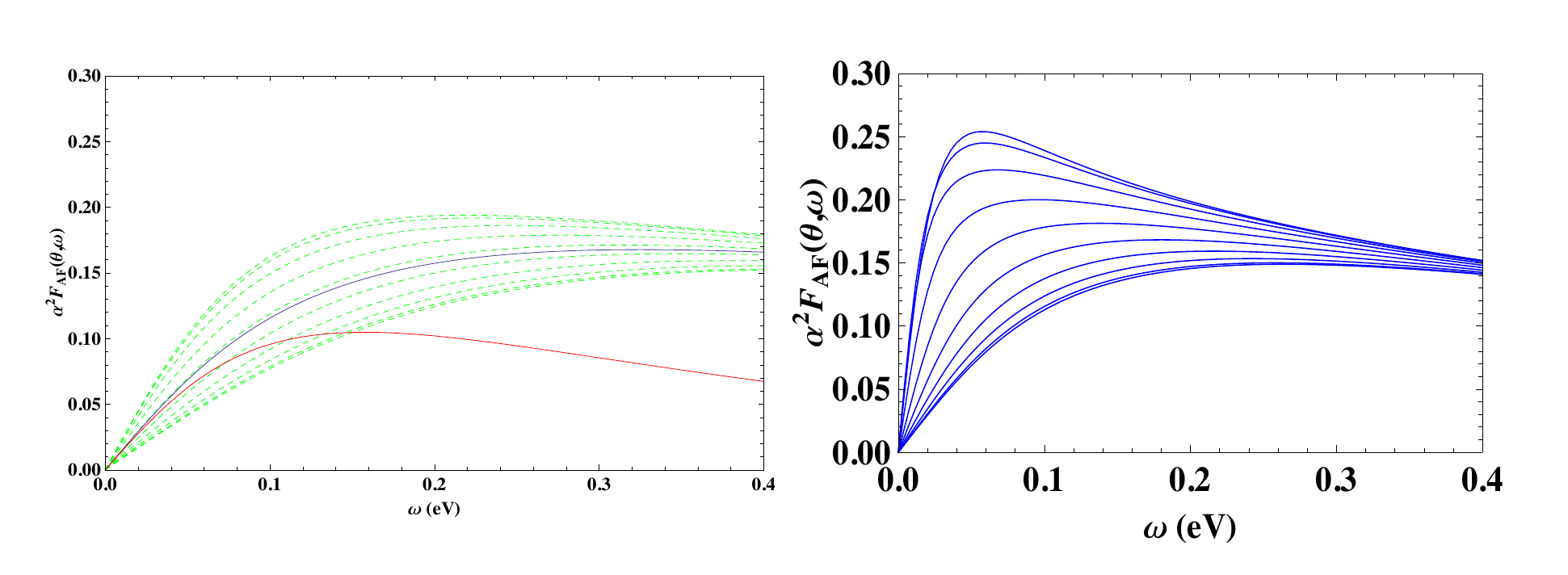}
\caption{The left panel in green shows the result for  $\theta$'s as in Fig. (\ref{AF1}) for $\xi/a =1$ and $\omega_{AF} =0.1 eV$. The red and the blue curves are again as defined in the caption for Fig. (\ref{AF1}). The $T_c$ calculated with these parameters for AFM fluctuations is 23 K. The right panel shows the variations for the same directions for $\xi/a =2$ and $\omega_{AF}$ =0.03 eV. The $T_c$ calculated for these parameters is 64 K.}
\label{AFx}
\end{figure}

 \noindent
$m_0^2$ fixes the integrated spectral weight of the fluctuations, $\xi$ is the correlation length and $\omega_{AF}$, the damping rate of the fluctuations. $(\alpha^2 F(\theta,\omega))_n$ can be obtained from Eq.(\ref{eq:chiaf}) after integrating over $\vk'$. We have followed this procedure to to fit the experimentally deduced curve as well as possible, fixing the overall magnitude with $m^2$ and adjusting the shape with the other two parameters. Results of $\xi/a$ = 1 and $1/\pi$ are shown in 
Fig. (\ref{AF1}). While the latter gives nearly momentum independent results, the former does so at higher energies but not at the lower energies. The discrepancy worsens on increasing $\xi/a$ or increasing $\omega_{AF}$, as shown in Fig. (\ref{AFx}). We then se Eq.(\ref{eq:chiaf}) with these parameters in the Eliashberg Equations (not their approximate solutions in terms of Eq. (\ref{lambdas})) to determine $T_c$. For $\xi/a = 1/\pi$, we find it to be below $0.1 K$ for all choices of $\omega_{AFM}$, for $\xi/a =1$, $T_c$ is calculated as 12 K for $\omega_{AF} = 0.03 eV$ and 27 K for $\omega_{AF} = 0.1 eV$. Only when $\xi/a$  is larger than 2, do we calculate  values of  $T_c$ of order $10^2 K$. The experimentally deduced  $(\alpha^2F(\theta, \omega))_n$ does not allow $\xi/a$ more than about $1/2$. Although no neutron scattering results in the normal state are available for Bi-2212 in the normal state, a correlation length $\xi/a \lesssim 1$ of less than a lattice constant is consistent with neutron scattering results in $YBa_2Cu_3O_{6+\delta}$ with $\delta$ near the highest $T_c$ \cite{mook-prl93}, \cite{bourges-balatsky}.

The conclusion from the above  analysis is that {\it if} the fluctuations coupling to fermions revealed in ARPES experiments were  AFM fluctuations, $T_c$ would have been less than 10 K. The reason for this is clear from looking at Fig (\ref{AF1}). If the correlation length of fluctuations is not much larger than the lattice constant, the projection to d-wave scattering, i.e. the calculation of the relative dominance of scattering of fermions through $\pi/2$ is small compared to the average or s-wave scattering. Hence $\lambda_d$ of Eq. (\ref{lambdas}) calculated as a (weighted) integral over the red curve in Fig. (\ref{AF1}) is much smaller than $\lambda_n$ which is calculated from the black curve. The same conclusions apply to  fluctuation spectra of any other operator which is of the form of (\ref{eq:chiaf}) because the data forces it to have a short correlation length. More sophisticated correlation functions than (\ref{eq:chiaf}) cannot change this conclusion significantly.

\subsection{Quantum-Critical spectra of Loop-Current Order}

We show in this section that a nearly $\theta$ independent $(\alpha^2F(\theta, \omega))_n$ as well as the $(\alpha^2F(\omega))_{p,d}$ to obtain high $T_c$ is consistent
with the spectral function of the local quantum-critical fluctuations derived recently  
 \cite{aji-cmv-qcf,aji-cmv-qcf-pr, aji-cmv-qcf-warps} due to the quantum melting of the loop-current order \cite{simon-cmv}, \cite{cmv-2006}
observed universally \cite{Fauque06prl,Mook08prb,Li08nature,Kaminski02nature}
in the underdoped region of the cuprates, and the coupling function of these fluctuations to fermions 
(\onlinecite{Aji-shekhter-cmv-prb-10}). We briefly summarize the results.

There are two aspects to the solution: the momentum independence  of the derived fluctuation spectra ${\mathcal F}({\bf q}, \omega)$ and the momentum dependence of the coupling function of such fluctuations to the fermions, $g({\bf k}, {\bf k+q})$.

The loop order is specified by a polar time-reversal odd vector ${\bf L}$ which specifies a pair of magnetic fluxes in each unit-cell generated through many body interactions. There are four possible orientations of ${\bf L}$. In the quantum-fluctuation regime, the orientations of ${\bf L}(i,t)$ in the cells $i$ may be taken to be a continuum. The fluctuations are specified  (\cite{aji-cmv-qcf}, \cite{aji-cmv-qcf-pr}, \cite{aji-cmv-qcf-warps} by the angular momentum operator $U(i,t)$ which quantum-mechanically flips local configurations of the loop order.  The Fourier transform of $<U^+(i,t)U(j,t)>$ in this theory are the propagator of the relevant fluctuations ${\mathcal F}({\bf q}, \omega)$ of Eq.(\ref{irr-int}). The spectral weight is locally critical and of the form,
\begin{eqnarray}
\label{eq:flucspec}
\Im {\mathcal F}({\bf q},\omega) &=& \begin{cases}
 -\chi_0 \tanh(\omega/2T), &|\omega| \lesssim \omega_c;  \\
0,  &|\omega| \gtrsim \omega_c.
\end{cases}
\end{eqnarray}
This is precisely of the form which leads to the marginal fermi-liquid properties \cite{mfl} in a part of the phase diagram. In the pseudogap region where loop order is observed through condensation of the low energy part of these fluctuations, this spectrum is depleted at low energies with some of the spectral weight going to weakly momentum dependent collective modes (\cite{he-varma-prl11}). Such modes have recently been observed \cite{li2}. 

 In the continuum limit, $U({\bf r})$ is the angular momentum operator generating rotations among the configurations of loop order.
Therefore it can only couple to the local angular momentum of fermions. So the coupling is of the form
\be
\label{coupl}
H_{int} \propto \int d{\bf r} \sum_{\sigma}g_0 ~ \psi^+({\bf r},\sigma) ({\bf \hat{r} \times \hat{p}}) \psi({\bf r},\sigma) {\bf U}({\bf r}) + H.C.
\ee 
$g_0$ is the coupling energy which can be estimated from experiments on single particle scattering rate. $H_{int}$ has also been explicitly derived for the fermions in a two-dimensional lattice model of the Cuprate lattice \cite{Aji-shekhter-cmv-prb-10}. It is instructive to note that Eq. (\ref{coupl}) is the orbital angular momentum analog of the familiar collective spin-fluctuation coupling to spin-flip excitations of fermions.

We may write Eq.(\ref{coupl}) in momentum space;
\be
\label{coup-kspace}
H_{int} = \sum_{{\bf k, k'}, \sigma} g_0 ~ i ({\bf \hat{k}} \times {\bf\hat{k'}})  \psi^+({\bf k},\sigma)\psi^+({\bf k'},\sigma) \cdot {\bf U}({\bf k-k'}) + H.C.
\ee

This identifies that in the continuum limit (similar conclusions are obtained from the lattice version of the calculations), the coupling function $g({\bf k}, {\bf k}')$ of Eq. (\ref{coupl}) is given by 

\be
\label{g-cont}
g({\bf k}, {\bf k}') = g_0 ~ i ({\bf \hat{k}} \times {\bf \hat{k}}'),
\ee
giving immediately that the strongest scattering by the fluctuations rotates the momentum of the fermions by $\pi/2$. As discussed in Sec. II this promotes "d-wave" pairing if the effective interaction is attractive even though the
 the spectrum of fluctuations itself is momentum independent.  That it is attractive may be seen as follows:
 Integrating over the fluctuations in Eq.(\ref{coup-kspace})gives an effective vertex for scattering of Cooper-pairs:

\begin{equation} \label{kxk'}
\Lambda\left(\textbf{k},\textbf{k},\right) \propto -({\bf k} \times {\bf k}')^2\Re {\mathcal F}({\bf{ k}}-{\bf{k'}},\omega).
\end{equation}
Since  $\Re {\mathcal F}({\bf k}-{\bf k'}),\omega) < 0$  for $-\omega_c <\omega < \omega_c$, independent of momentum, the pairing symmetry is given simply by expressing $({\bf k} \times {\bf
k}') ^2$ in separable form~:
\begin{align} ({\bf k} \times {\bf k}') ^2 &= 1/2 \left[(k_x^2+k_y^2)(k_x^{'2}+k_y^{'2})
-  (k_x^2-k_y^2)(k_x^{'2}-k_y^{'2})\right.\nonumber \\
&- \left. (2 k_xk_y) (2 k'_x k'_y)\right].
\end{align}
Pairing interaction in  the $s$-wave channel is repulsive, that in the two $d$-wave channels is equally attractive, and in the odd-parity channels is zero. The factor $i$ in $g({\bf k}, {\bf k}')$, present because the coupling is to fluctuations of 
time-reversal odd operators, is crucial in determining the sign of the interactions of the pairing vertex. In the lattice version of the theory \cite{Aji-shekhter-cmv-prb-10}, the weighting of the angular decomposition by the density of states in different irreducible representations favors the $(x^2-y^2)$ pairing over the $2 x y$ pairing.

\begin{figure}
\includegraphics[scale=0.7]{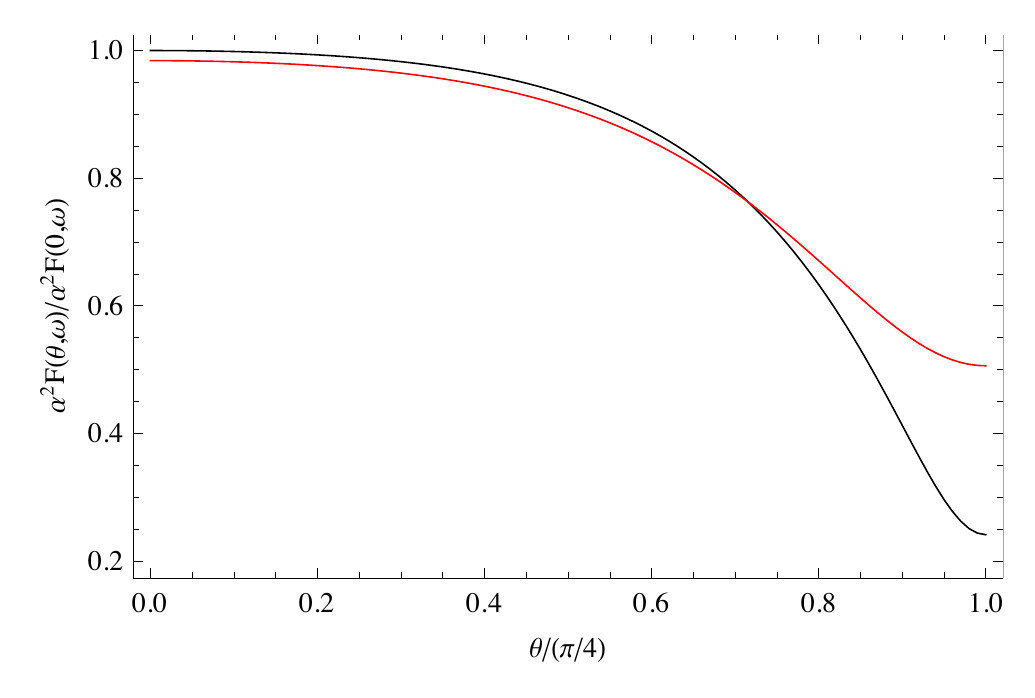}
\caption{The angle dependence for any frequency expected for $\alpha^2F_n(\theta, \omega)$ in the coupling of loop-order fluctuations to fermions in a simple model for the matrix elements. The red-curve is for the ratio of velocity in the ($\pi,\pi$) direction to that in the ($\pi,0$) direction of 2 and the black curve for 4. The red-curve is slightly displaced with respect to the black curve for clarity.}
\label{alpha2(theta)}
\end{figure}

The resolution of the puzzle of a nearly $\theta$ independent $\alpha^2F_n(\theta, \omega)$ in Fig.(\ref{normal F}) with a strong pairing in the d-wave channel comes about in the following way. Consider first the part of Fig.(\ref{normal F}) without the bump. This frequency dependence (including the angle dependent cut-off) automatically comes from the derived spectra of Eq. (\ref{eq:flucspec}) and this gives no $\theta$ dependence in $\alpha^2F_n(\theta, \omega)$, which comes only from integrating 
$|g({\bf k}, {\bf k}')|^2$ over ${\bf k}'$ as in Eq. (\ref{a2Fn}) and then converting the resulting function of ${\bf k}$ to that of $\theta$ by multiplying with the local density of states at the chemical potential $N(\theta)$. In the continuum limit Eq.(\ref{g-cont}), this automatically yields a $\theta$-independent result. We have also calculated it from the leading  lattice-dependent $g({\bf k}, {\bf k}')$, given in Ref.\cite{Aji-shekhter-cmv-prb-10}:
\bea
g({\bf k}, {\bf k}') \propto i g_0 \big(\sin(k_xa) \sin(k_y'a) - \sin(k_ya) (k_x'a)\big)
\eea
The results are shown in Fig. (\ref{alpha2(theta)}) for two choices of $g_0$, which determines the ratio of the fermi-velocity in the ($\pi,\pi$) directions to that in the ($\pi,0$) directions to be 2 and 4. The  experimental result for this ratio is about 2. The variation in the $\theta$ dependence in the measured range, $\pi,\pi$ direction to about $\pi/8$ from it is about 10\%, which on comparison with Fig. (\ref{normal F}) is within the experimental uncertainty. With higher resolution data, the prediction is that there should be a frequency-independent $\theta$-dependence of the order of magnitude given in Fig. (\ref{alpha2(theta)}).
.

There is one aspect of
the deduced $\alpha^2F(\theta, \omega)$ which is not given by the critical
theory. This is the low energy bump as a function of $\omega$.  If the bump occurs 
 only in samples studied in the pseudogap region, it is probably due to the collective modes which have been discovered to be
 special to this region of the phase diagram \cite{li2, he-cmv}. This can be checked by equally high
resolution data in samples in the quantum-critical region or the overdoped region of the
phase diagram. Alternatively, it may be due to coupling to oxygen vibrations modes, as has been suggested \cite{shen-phonons}.

In the loop-fluctuation theory, the equivalent of the red and the black curves of Fig. (\ref{AF1}) used in the Eliashberg Equations to calculate $T_c$ have identical frequency dependence and nearly the same magnitude. The calculated $T_c$ using the experimentally deduced $\alpha^2F(\theta, \omega)$ is about 100 K. There remain valid questions about the applicability of the Eliashberg equation to spectra whose cut-off is comparable to the over-all electronic bandwidth and with coupling constants of about $1/2$. We believe that the best way to investigate this is to deduce the variations in $\alpha^2F(\theta, \omega)$ for a variety of dopings and $T_c$ in a given compound. We doubt that Eliashberg theory works quantitatively to the accuracy that it does for electron-phonon interaction but we expect to get the ratios of calculated $T_c$'s for different dopings to compare well with the ratio of observed $T_c$'s. Such work is in progress. 

\section{Concluding Remarks}

  \begin{figure}[htbp]
\begin{center}
\begin{minipage}{1.0\textwidth}
\includegraphics[width=1.0\textwidth]{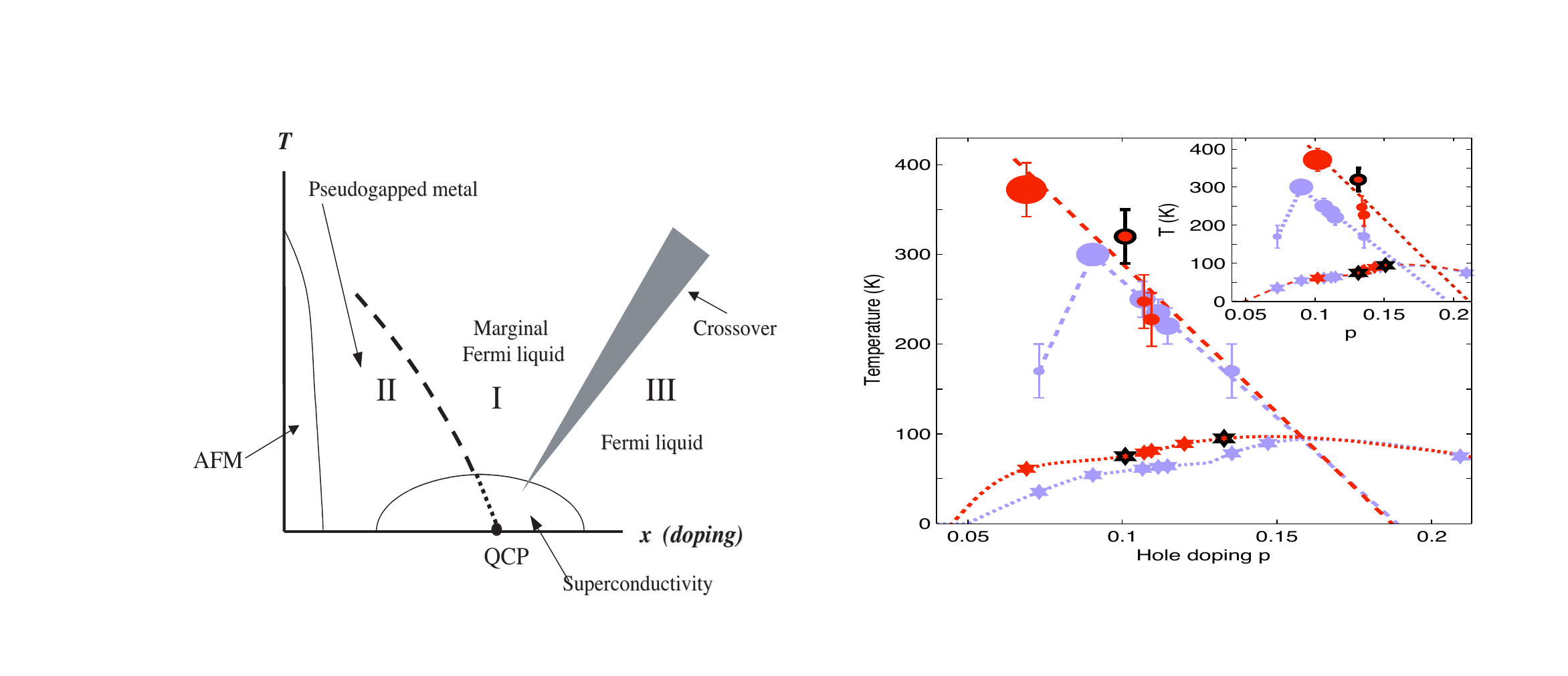}
\caption{The figure on the left shows the proposed \cite{cmv-prb97} Universal Phase diagram for hole-doped cuprates.   The right part of the figure (taken from Ref.(\cite{li-prb})
 shows the lines of phase transitions from experiments in Hg1201(red) and YBCO(blue) at low and intermediate dopings. The main panel and the inset show data plotted with two different way of determining $p$. Stars denote $T_c$, and circles denote $T^*$, the temperature below which the predicted magnetic order is seen.}
\label{phase-dia}
\end{minipage}
\end{center}
\end{figure}

The phase diagram in Fig. (\ref{phase-dia}) may be taken to provide a motivating principle in the search for a theory of the Curpates - that the universal properties of all the regions in it be governed by a single idea which can be calculated in a controlled fashion for properties in all the regions and with predictions which are testable in experiments.

The left part of Fig. (\ref{phase-dia}) divides the $T-doping$ plane of hole-doped cuprates into regions with different characteristic  thermodynamic and transport properties. Such a diagram was first proposed in 1997 \cite{cmv-prb97, tallon}, but only in the past five years have some of its principle features been verified in experiments. The essential feature of the phase diagram is a quantum-critical point from which emanate three regions - a region II of broken symmetry, a region I whose properties are determined by the quantum-critical fluctuations of a proposed order in region II, with a crossover to a fermi-liquid region III. D-wave superconductivity straddles the three regions at low temperatures. 
A time-reversal breaking order, preserving translational symmetry was predicted for the region II. Such an order has been observed by polarized neutron scattering in $YBa_2Cu_3O_{6+x}$ \cite{fauque} , HgBaCuO \cite{li-elastic}, LaSrCuO \cite{mesot-lsco}, BISCCO \cite{bourges-pc}, the last consistent with earlier dichroic ARPES experiments \cite{kaminski}. The second part of Fig. \ref{phase-dia} taken from Ref.(\cite{li-prb}) shows the results for $T_c$ and the temperature of the predicted order, which coincides within uncertainty with $T^*$ determined by other methods for the first two compounds.  In the underdoped region, other orders have been discovered; they are different in different cuprates and do not demarcate the universal change in properties as the lines drawn in Fig. (\ref{phase-dia}) do. For example the sharp decrease in the order in $YBa_2Cu_3O_{6+x}$ at small x coincides with the onset \cite{Bourges-Sidis} possibly of {\it smectic} or stripe order.

 The motivation for proposing a symmetry breaking in the pseudogap regime was the "strange-metal" Region I, whose properties could be understood by hypothesizing a quantum-critical spectrum \cite{mfl}. Two major theoretical developments since then have been the microscopic derivation of this quantum critical spectrum as due to the fluctuations of the order parameter in Region II, and the derivation of the matrix element for coupling the fluctuations to fermions and the proof that it promotes d-wave pairing as discussed above. 

The different aspects of the theory: the loop-current order, its quantum-critical fluctuations and their coupling to fermions are all essential and used in the theory of d-wave superconductivity summarized above.  We have shown that the detailed analysis of high resolution ARPES yields results which are very specific in their angle and frequency dependence and that they are consistent for $T_c$ in d-wave symmetry from the same theory with which aspects of the normal state in Region I are understood and with which the order observed in Region II was predicted. There are aspects of experiments in Region II, such as the single-particle spectra and the magneto-oscillations which however remain to be understood on the basis of this (or any other) ideas.

We have also shown that any existing theory based on momentum-dependent fluctuation spectra such as AFM fluctuations would need to have a correlation length smaller than the lattice constant to be consistent with the deductions from ARPES and that such small correlation lengths would yield $T_c$ in the d-wave channel with too low a value. This is hardly surprising since theories based on such ideas have not succeeded in understanding the normal state properties in the Cuprates.

{\it Acknowledgements}: Thanks are due to Elihu Abrahams for a critical reading of the manuscript. CMV's research is supported by a grant from the Division of Materials Research of the National Science Foundation.

\bibliography{at}
\bibliographystyle{unsrt}

\end{document}